# A big data intelligence marketplace and secure analytics experimentation platform for the aviation industry

Dimitrios Miltiadou[1][0000-0001-9030-3299], Stamatis Pitsios[1][0000-0001-9525-1416], Dimitrios Spyropoulos[1][0000-0002-5820-4319], Dimitrios Alexandrou[1][0000-0002-2712-2089], Fenareti Lampathaki[2][0000-0002-3131-4622], Domenico Messina[3][0000-0001-7455-2623], Konstantinos Perakis[1][0000-0002-2236-9816]

[1] UBITECH, Thessalias 8 & Etolias, Chalandri, 15231, Greece
[2] SUITE5, Alexandreias 2, Bridge Tower, Limassol, 3013, Cyprus
[3] ENGINEERING Ingegneria Informatica S.p.A., Piazzale dell Agricoltura 24, Rome, 00144, Italy

**Abstract.** Over the last years, the impacts of the evolution of information integration, increased automation and new forms of information management are also evident in the aviation industry that is disrupted also by the latest advances in sensor technologies, IoT devices and cyber-physical systems and their adoption in aircrafts and other aviation-related products or services. The unprecedented volume, diversity and richness of aviation data that can be acquired, generated, stored, and managed provides unique capabilities for the aviation-related industries and pertains value that remains to be unlocked with the adoption of the innovative Big Data Analytics technologies. The big data technologies are focused on the data acquisition, the data storage and the data analytics phases of the big data lifecycle by employing a series of innovative techniques and tools that are constantly evolving with additional sophisticated features, while also new techniques and tools are frequently introduced as a result of the undergoing research activities. Nevertheless, despite the large efforts and investments on research and innovation, the Big Data technologies introduce also a number of challenges to its adopters. Besides the effective storage and access to the underlying big data, efficient data integration and data interoperability should be considered, while at the same time multiple data sources should be effectively combined by performing data exchange and data sharing between the different stakeholders that own the respective data. However, this reveals additional challenges related to the crucial preservation of the information security of the collected data, the trusted and secure data exchange and data sharing, as well as the robust access control on top of these data. The current paper aims to introduce the ICARUS big data-enabled platform that aims provide a multi-sided platform that offers a novel aviation data and intelligence marketplace accompanied by a trusted and secure "sandboxed" analytics workspace. It holistically handles the complete big data lifecycle from the data collection, data curation and data exploration to the data integration and data analysis of data originating from heterogeneous data sources with different velocity, variety and volume in a trusted and secure manner.

## 1    Introduction

The Aviation industry encompasses a wide range of activities and industries directly linked to aircrafts' development, production and operation, as well as an extensive list of interrelated products and services that support the aircrafts' overall operations. Despite the fact that the aviation industry has successfully been raised into a leading and critical industry of the global economy, the various sectors of the industry operate on a fragmented manner. Over the last years, the impacts of the evolution of information integration, increased automation and new forms of information management are also evident in the aviation industry that is disrupted also by the latest advances in sensor technologies, IoT devices and cyber-physical systems and their adoption in aircrafts and other aviation-related products or services. It is also now acknowledged in the aviation industry that the aircraft owners have now huge amount of data and information about their aircrafts, which are necessary and have real value impact on their aircrafts so they are in need to maximise the use of technology [1]. This is also obvious from the latest estimations reporting the generation between 500 and 1,000 gigabytes of data in on an average flight level [2], while on global fleet level the generation of up to 98,000,000 terabytes of data by 2026 [3]. To this direction, the unprecedented volume, diversity and richness of aviation data that can be acquired, generated, stored, and managed provides unique capabilities for the aviation-related industries and pertains value that remains to be unlocked with the adoption of the innovative Big Data Analytics technologies.

In the Big Data era a tremendous amount of information is generated in an increasing immeasurably magnitude from of a plethora of sources that should be effectively collected and harnessed in order to be properly processed towards the extraction of valuable knowledge and added value, the reveal of trends and hidden patterns or correlations that will facilitate the prediction and decision making[4][5]. To accomplish this, a new generation of technologies that are referred as Big Data technologies have been arising capable of extracting added value from an enormous volume of data with rich diversity towards the effective and efficient high-velocity capture, discovery, and analysis [6]. In this sense, the big data technologies are focused on the data acquisition, the data storage and the data analytics phases of the big data lifecycle by employing a series of innovative techniques and tools that are constantly evolving with additional sophisticated features, while also new techniques and tools are frequently introduced as a result of the undergoing research activities. Nevertheless, despite the large efforts and investments on research and innovation, the Big Data technologies introduce also a number of challenges to its adopters. Besides ensuring the employment of effective storage and access mechanisms, the crucial aspects of offering efficient mechanisms that enable data integration and data interoperability should be considered. Furthermore, in order to be able to execute data analytics that would generate valuable insights and information it is imperative that multiple data sources should be effectively combined. As a consequence, this usually implies that exchange and sharing of data is performed between the different stakeholders that own the respective data. However, this reveals additional challenges related to the crucial preservation of the information security of the collected data, the trusted and secure data exchange and data sharing, as well as the robust access control on top of these data.

The current paper aims to introduce the ICARUS big data-enabled platform that aims provide a multi-sided platform that offers a novel aviation data and intelligence marketplace accompanied by a trusted and secure "sandboxed" analytics workspace. To this end, the platform holistically handles the complete big data lifecycle from the data collection, data curation and data exploration to the data integration and data analysis of data originating from heterogeneous data sources with different velocity, variety and volume in a trusted and secure manner. The platform exploits methods such as big data analytics, deep learning, data enrichment, and blockchain powered data sharing, in order to properly address critical barriers for the adoption of Big Data in the aviation industry facilitating the design and execution of big data scenarios from the stakeholders of the aviation industry.

## 2   Materials and Methods

Despite the embracement of the big data technologies in different domains and their advancements, it is proven that several challenges have not been yet addressed since the majority of the described big data technologies are still in their early stages. It is acknowledged that the crucial challenge in any big data platform is to gather, store, search, share, transfer, analyse and present data while ensuring that compliance to the identified requirements is maintained [7]. At the moment, the efforts are focused on effectively storing and accessing large data that are originating from heterogeneous and diverse data sources and are created in multiple (structured, semi-structured, unstructured) formats, while other aspects such as the data integration and data interoperability are often neglected. Furthermore, while big data analytics tools are constantly evolving and empowered with new features, less effort is spent in supporting seamless and effortless analytics on top of cross-origin data by the developed analytics tools. In the same context, the effective handling of information security while storing and managing of this vast amount of data from heterogeneous data sources remains a crucial challenge. The extraction of valuable knowledge and intelligence unavoidably requires the dynamic data exchange and data sharing between the various stakeholders of an industry or across different industries in a trusted, regulated and secure manner. Furthermore, data access control on top of these collected massive and rapidly evolving data must be properly addressed. Hence, it is also acknowledged that despite the constantly growing number of available technologies and techniques that have emerge, there is a real challenge on finding the proper balance between the effectiveness and performance of the dynamic analysis on diverse large data sets and the requirements for data integrity, data governance and data security and privacy [8].

Nevertheless, a promising opportunity arises from the latest developments and compelling features of the big data technologies to design and build a big data platform that capitalizes on these emerging offerings in order to build a novel data value chain in the aviation-related sectors that will enable data-driven innovation and collaboration across currently diversified and fragmented industry players, by effectively addressing the challenges imposed by the nature of big data and the requirements of the aviation industry's stakeholders for a trusted and secure data sharing and data analysis environment.

### 2.1 The ICARUS technical solution

With a view to fulfil these objectives and address the challenges mentioned in the previous section, a thorough analysis of the collected requirements from the aviation stakeholders was performed towards the design and development of a big data-enabled platform that aspires to provide an intuitive aviation data and intelligence marketplace that provides a trusted and secure "sandboxed" analytics workspace.

The main objectives of the ICAURS platform [9] is to provide an innovative and intelligent big data-enabled environment that enables the aviation sector's data providers and data consumers to effectively and securely the complete big data lifecycle that includes initially the data preparation and upload as a first step, the data exploration, sharing and brokerage as a second step towards the final step of the data analysis and data visualisation. In this context, the platform incorporates at its core advanced data management and data value enrichment methods that span over the axes of data collection, data curation, data safeguarding, data sharing and data analytics in order to effectively and efficiently cover all the aviation industry's needs, requirements and peculiarities with regards to big data, as well as to knowledge and insight extractions from these data.

To this end, we are developing an integrated big data-enabled platform that is composed by a set of key components that are designed and implemented by exploiting well-established and state-of-the-art big data infrastructure, technologies and tools. The architecture of the platform is a **modular architecture**, composed by **22 components** in total, that is designed with the aim to offer the maximum flexibility and extensibility, enabling the smooth integration and effective operation of the various components that are implemented as distinct software modules. The designed software modules are exploiting and combining multiple technologies and tools towards the aim of accomplishing the aspired offerings of the platform.

During the design process, the major focus was on the functional decomposition, the strict separation of concerns, the dependencies identification and especially the data flow realisation. To this end, each component has been designed in order to operate under a clear context, with distinct features and functionalities and a clearly defined scope within the architecture. The elicited technical requirements and functional specifications were carefully analysed and facilitated the evolution of a mature concept architecture design that is aiming to address the ambition of delivering a novel big data platform for the aviation data value chain.

Towards this aim, the design of the platform ensures the offering of a scalable and flexible environment that will enable the interoperability of the various components that facilitate the execution of big data analytics and sharing of data through secure, transparent and advanced functionalities and features. The designed architecture incorporates all the entire lifecycle of the platform that spans from data preparation and data upload, to data exploration, data sharing, data brokerage and data analysis. To cope with the security and privacy constraints of the aviation sector, the platform adopts a security and privacy by-design approach that covers all the data confidentially, data privacy and data safeguarding aspects of the platform. The main axes of this approach is the end-to-end encryption of all the datasets that are available in the platform with a symmetric encryption key and the design and employment of a sophisticated secure decryption process with a decryption key in the form of key pair for each dataset per data consumer to facilitate the secure data sharing of the datasets across all the platform

The platform architecture is conceptually divided in three main tiers, namely the **On-Premise Environment**, the **Core Platform** and the **Secure and Private Space**, and each tier is composed by a set of components, from the total list of 22 components, that are combined towards the realization of the functionalities of each specific tier in the underlying execution environment (see Fig.1).

The On-Premise Environment is responsible for the execution of the data preparation and data uploading functionalities and features of the ICARUS platform based on the instructions that are provided by the Core Platform in accordance with the preferences of the data provider. The On-Premise Environment is composed by multiple components that are running on the data provider's environment which are utilised towards the preparation and uploading of the data provider's private or confidential datasets to the Core Platform.

The Core Platform constitutes the main tier of the platform which undertakes the execution of all the core operations of the platform and the formulation and propagation of instructions to the On-Premise Environment and the Secure and Private Space tier for local execution. In this context, in the Core Platform multiple components are integrated and combined in the platform's cloud infrastructure in order to execute all the data exploration, sharing and brokerage operations, as well as the design of the data preparation, data uploading and data analysis operations that will be executed by the On-Premise Environment and the Secure and Private Space respectively. Furthermore, the Core Platform is offering the user interface of the platform that users utilise to perform all these operations.

The Secure and Private Space constitutes the advanced analytics execution environment of the platform in which all the data analytics processing is executed in an isolated, secure and trusted way. The Secure and Private Space is composed by a set of components that formulate a trusted and secure sandboxed analytics workspace in which the data analysis executed in accordance with the analytics workflow that is designed by the user within the Core Platform. Hence, the Secure and Private Space is offering a rich stack of analytics tools and features whose management and orchestration is performed through the Core Platform and whose operation is performed by adhering the strict and rigorous security and privacy needs of the aviation sector.

In the following paragraphs the three tiers of the ICARUS platform are presented, focusing in the components that compose each tier as well as the interactions between them.

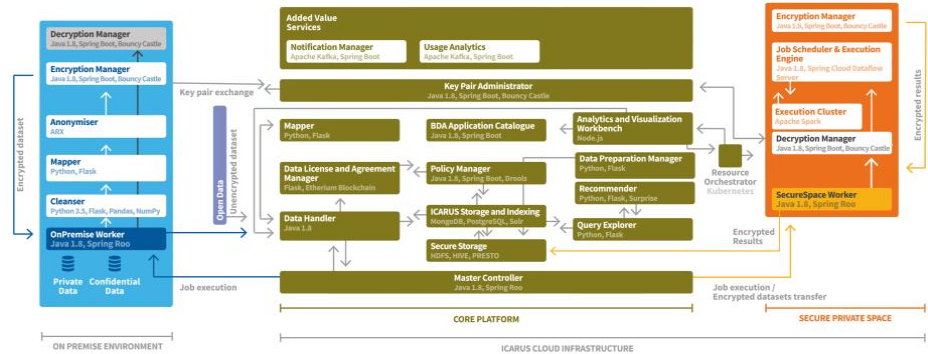

**Fig. 1.** The ICARUS platform conceptual architecture

**On-Premise Environment**
. The scope of the On-Premise Environment is to provide the required services that will perform all the data preparation and data uploading steps as instructed by the Core Platform and is composed by the On-Premise Worker, the Cleanser, the Mapper, the Anonymiser, the Encryption Manager and the Decryption Manager.

The scope of the **On-Premise Worker** is to execute the jobs or tasks, according to the instructions provided by the Master Controller component of the Core platform, utilising the available components on the On-Premise Environment. Hence, the On-Premise Worker interprets and executes the instructions for the job or task execution and provides the execution status of the requested jobs/tasks back to the Master Controller. Finally it supports the uploading of the prepared encrypted datasets from the On-Premise Environment to the Core Platform, as well as the downloading of the datasets to the local environment via the interaction with Data Handler component of the Core Platform. The On-Premise Worker provides the user interface of the On-Premise Environment through which the user is able to perform all the described operations.

The **Cleanser** is the component undertaking the responsibility of the data cleansing functionalities of the platform. The Cleanser supports a set of techniques for performing simple and more advanced cleansing operations over datasets that contain erroneous or "dirty" data by detecting or correcting corrupted, incomplete, incorrect and inaccurate records from datasets with a variety of rules and constraints. For this purpose, the Cleanser employs a set of validation, cleansing and missing value handling rules based on the user input. Additionally, it provides the logging mechanism that monitors and stores all the identified errors, the actions performed and the corresponding results.

The **Mapper** is responsible for the harmonisation process of the dataset by enabling the user to define the mapping of the fields of the dataset to the common aviation model employed in the platform in a semi-automatic way. Moreover, the Mapper enables the exploration of the common aviation model from the user in order to provide suggestions for possible extensions of the model. The Mapper has a dual presence in the architecture: a) it provides a graphical user interface as part of the Core Platform that facilitates the definition of the mapping between the dataset fields and the common aviation model and b) a backend mapping service operating on the On-Premise Environment that executes the actual defined mapping from the user's data to the common data model based on the user's input from the user interface. In a nutshell, the Mapper generates the proposed mapping from a new dataset to the common data model and allows the users to review and update it based on their knowledge and expertise on their datasets. The defined mapping is provided in the form of an instructions to the backend mapping service that resides in the On-Premise Environment through the Master Controller and the On-Premise Worker. The defined mapping can be also stored and reused as mapping template.

The **Anonymiser** is providing the data anonymisation functionalities in order to filter or hide the private, sensitive or personal data that cannot be disclosed outside the data provider's premises, corporate network or personal filesystem by providing the means to deal with privacy issues and protection of sensitive information with a variety of anonymisation techniques. It employs a privacy and anonymisation toolset with a variety of privacy models such as the K-anonymity, the L-diversity and T-closeness models and anonymisation techniques such as the Generalisation, Masking, Micro-aggregation and Clustering and Micro-aggregation in order to handle the incoming datasets and provide the anonymised datasets.

The **Encryption Manager** is undertaking all essential encryption processes with regard to encryption of the data provider's dataset. It provides the encryption cipher mechanism that generates the symmetric encryption key and the encrypted dataset by performing on-demand column-based encryption of a dataset. Moreover, it facilitates the dataset sharing, upon the agreement of the data provider and the data consumer, with the generation of the appropriate decryption keys, in the form of key pairs, one per dataset per data consumer. It undertakes the storage and management of the generated decryption keys and the secure transmission of the corresponding decryption key from the data provider to the data consumer. Finally, it maintains a local key store where generated decryption keys and all relevant information is stored and handles the revocation requests in case access to an encrypted dataset is revoked for a specific data consumer.

The **Decryption Manager** is enabling the decryption of the dataset on the On-Premise Environment when an encrypted dataset is downloaded locally, provided that a valid smart contract, handled by the Data License and Agreement Manager of the Core Platform, exists permitting the downloading of the specific dataset locally. The Decryption Manager provides the mechanisms to verify the identity of the data consumer via a certificate or public key, to request for the decryption key from the data provider and the decryption mechanism in order to temporarily reproduce the encryption key in order to decrypt the dataset. It interacts with the Key Pair Administrator residing on the Core Platform to initiate the request for a dataset that will trigger the generation of the decryption key from the data provider and facilitates the establishment of the secure connection with the data provider for decryption key transmission.

**Core Platform**

The scope of the Core Platform is to provide all the required components for the execution of the core operations of the platform, as well as compilation of the instructions that are executed by the On-Premise Environment and the Secure and Private Space. The Core Platform is composed by the core data components, namely the Master Controller, the Data Handler, the Data License and Agreement Manager, the Policy Manager, the Storage and Indexing, the Query Explorer, the Recommender, the Key Pair Administrator, the Resource Orchestrator, the Secure Storage, the Data Preparation Manager, the Analytics and Visualisation Workbench and the BDA Application Catalogue, as well as the added value components, namely the Notification Manager and the Usage Analytics.

The **Master Controller** is the component responsible for compiling a set of instructions for the execution of specific jobs or tasks, as provided by the components of the Core platform, and for providing this set of instructions for local execution to the workers running on the On-Premise Environment and the Secure and Private Space, namely the On-Premise Worker and the Secure-Space Worker. Additionally it monitors the execution status of requested jobs or tasks. Furthermore, it transfer the list of selected encrypted datasets to the Secure Storage and supports the uploading of the encrypted analysis results from the Secure and Private Space to the Secure Storage.

The **Data Handler** component encapsulates various services responsible for tasks related to making data available from and to the platform, as well as among different platform components. It serves as the "data gateway" in the architecture, as it supports the complete workflow of uploading proprietary and open datasets to the platform in order to be stored in the platform's storage, downloading datasets from the platform to the end user's On-Premise Environment and/or to the Secure Storage in order to be utilised by the Secure and Private Space, and finally the uploading of data generated in a Secure and Private Space back into the Secure Storage. The Data Handler interacts with the Master Controller in order to perform the data upload or download operations.

The **Data License and Agreement Manager** is the component responsible for handling all processes related to the data licenses and IPR attributes, as well as the drafting, signing, and enforcing the smart data contracts that correspond to data sharing agreements between platform users. It allows the users to define IPR related attributes, pricing terms and policies, as well as the license for the datasets they own in the platform. Furthermore, it implements the blockchain functionalities of the platform which hosts a local blockchain node and enables the users to create, edit, review, negotiate, update and decline or sign data sharing agreements that are translated into smart data contracts in the blockchain. Finally, it provides, checks or updates the status of a smart data contract.

The **Policy Manager** is the component providing the authorisation engine that implements the access control mechanisms within the platform. The purpose of the Policy Manager is to provide the logical access control that prevents the unauthorised access of any type of resource of the platform such as data, services, tools, any kind of system resources, as well as all other relevant objects. Hence, the Policy Manager is responsible for the implementation of the authorisation engine that is be based on the ABAC model and incorporates the required authorisation XACML-based policies. The purpose of this authorisation engine is the provide the access control decision that will either grant or deny the access to the requestor by enforcing the formulated authorisation policies. Additionally, the Policy Manager enables the definition, storage, reuse, update and disposal of the authorisation policies in order to allow the data providers to effectively define and manage the protection and sharing aspects of their datasets.

The **Storage and Indexing** is the component that enables the storage and indexing capabilities of the platform. This component is responsible for the effective and efficient storage and maintenance of large, complex and unrelated datasets within the platform, as well as the flexible and high-performance indexing of the stored datasets. It handles all the requests for storage or data retrieval from the various components of the platform, as well as the indexing of the incoming datasets, simple or geospatial query execution and full-text search on top of the stored datasets.

The **Query Explorer** is the component that offers dataset exploration and discoverability functionalities to the platform users. Query Explorer has two main offerings: (a) a graphical interface for users to search for datasets and view the search results and (b) a service that translates each search to a query that can be processed by the storage and indexing component. As such, the Query Explorer constitutes the main facilitator of the data marketplace functionalities and the main interaction point from the user's perspective. It enables the users to define search criteria, which includes selecting fields from the data model, define and apply filters based on the metadata or the actual unencrypted data of the datasets in order to discover potentially interesting datasets in the platform. Furthermore, it stores the query history of each user in order to be re-executed and provides dataset suggestions to the user upon interacting with the Recommender component.

The **Recommender** is providing the enhanced recommendation functionalities that enable the dataset exploration and discoverability. In particular, this component is responsible to recommend and suggest to the users additional related datasets during the search and query process using a combination of collaborative filtering and content-based filtering. Furthermore, the Recommender will provide recommendations based on the users' preferences and history of searches, requests and purchases.

The **Key-Pair Administrator** is the component that is facilitating the exchange of the decryption keys between the data consumer and the data provider in order to enable the end-to-end data encryption and secure sharing of the encrypted datasets. In this context, the Key-Pair Administrator performs the signalling operations between the data consumer and the data provider in order to achieve the establishment of a secure connection between these parties, by performing the identity verification of each party and the successful handshake between them. Moreover, it maintains the list of data providers that have provided encrypted datasets so as to facilitate the key exchange process, as well as the mechanism for the revocation of the decryption keys upon needs.

The **Resource Orchestrator** is enabling the provisioning and management of the Secure and Private Space in the form of a scalable isolated environment running on virtualised infrastructure. More specifically, the Resource Orchestrator is exploiting the concept of isolated containerised execution environments with easy monitoring, autoscaling and orchestration of applications utilising the Docker container technology. The Resource Orchestrator is utilising the well-established open source platform of Kubernetes for container orchestration that facilitates the deployment of the Secure and Private Space components in the form of containerized applications with extensive support of features like service discovery, load balancing, storage orchestration, automated rollouts and rollbacks, health monitoring and self-healing processes, while at the same time ensuring the required isolation and security features of the Secure and Private Space tier.

The **Secure Storage** is providing the secure and trusted storage of the Secure and Private Space with a two-fold purpose: i) to enable the effective and secure storage of the data provider's datasets, that includes the owned private and confidential datasets as well as the ones purchased through the marketplace of the platform, in order to be exploited by the Secure and Private Space in the context of data analysis and b) to provide the trusted and secure storage for the produced results of the executed data analysis. To its goals, the Secure Storage is formulated by a stack of big data-enabled state-of-the-art technologies towards the assurance of high availability and high performance for the computationally-intense data analysis operations with: a) the Hadoop Distributed File System (HDFS) that is provisioned and monitored by the Apache Ambari in the first layer for actual storage of both the datasets and the analysis results, b) the Apache Hive operating on top of the HDFS as a second layer, acting as a data warehouse software, offering an abstraction layer on top of the datasets residing in HDFS, while at the same time enabling the effective data lifecycle management of the underlying datasets and results with data query and analysis functionalities and c) the Presto high-performance query engine that offers the required fast parallel analytics query support against data of any size that are residing in Hive and consequently HDFS.

The **Data Preparation Manager** is enabling the data manipulation and data preparation operations on top of the datasets or results residing on the Secure Storage in order to be utilised by the Analytics and Visualisation Workbench. In this sense, the Data Preparation Manager facilitates the user to employ multiple data preparation operations in the form of sequential steps on his/her datasets in accordance to his/her needs for the data analysis or data visualisation. The data preparation operations spans from column creation (timestamp-related, math-related, aggregation related, shift and conditional operations), column drop and column renaming, row filtering to dataset merging, data completion and compute aggregations operations. Under the hood, the Data Preparation Manager is exploiting the Presto functionalities in order to create new versions of the dataset(s) within the Secure Storage in order to address the needs of the user for data analysis preparation.

The **Analytics and Visualisation Workbench** is providing the graphical environment where the users of the platform are able to design, execute and monitor the data analytics workflows and also where the visualisation and dashboards are displayed. The Analytics and Visualisation Workbench offers a large list of analytics algorithms and their customisation with a list of options per algorithm. Furthermore, it offers a variety of visualisation types that can be customised based on the user's needs. The Analytics and Visualisation Workbench decouples the design of the data analytics workflow from its actual execution. It offers a novel interactive canvas-based graphical user interface to the users in order to compose their desired data analytics workflows in the Core Platform while orchestrating and managing the actual execution of the analysis is that performed within the Secure and Private Space with the use of the Master Controller, the Secure Storage and the Secure-Space Worker that utilises the rest of the services of the Secure and Private Space in order to perform the actual data analysis execution. Furthermore, through the Analytics and Visualisation Workbench the advanced visualisation capabilities of the platform are offered through a modern data visualisation suite of charts and visualisations that span from basic charts to advanced multilevel visualisations which can be combined in order to form dynamic dashboards upon the user needs. In this context, the user is able to create an application, which contains the list of datasets that were selected for analysis, the selected algorithm, as well as the selected visualisation type, along with the corresponding parameters, and store it in the BDA Application Catalogue for later reuse.

The **BDA Application Catalogue** implements a repository of the applications created by the users of the platform. As such, the applications can be stored, retrieved, modified and loaded in the Analytics and Visualisation Workbench by the users at any time. The purpose of the BDA Application Catalogue is to enable the reuse of the designed data analytics workflows from the users, as well as the sharing of these workflows among the users through a defined license in order to further empower the analytical capabilities of the platform.

The **Notifications Manager** is responsible for providing the updated information to the users with regards to the datasets or the scheduled analytics jobs. More specifically, the Notifications Manager provides notifications to the users related to the availability of new datasets according to their configured preferences or any possible updates on the datasets that the users are entitled to use, as well as any updates on the execution status of the scheduled analytics jobs.

The **Usage Analytics** component is responsible for providing the tools that collect, analyse and visualise the usage of the various services and assets of the platform in order to extract useful insights and statistics. The Usage Analytics monitor the user's behaviour in various levels such as the usage and adoption of specific features or services and the usage of each dataset or algorithm towards the aim of providing usage information to both the users and the platform administrator.

**Secure and Private Space**

The scope of the Secure and Private Space is to provide all the required components for the formulation of the trusted and secure advanced analytics execution environment of the platform. The Secure and Private Space is composed by the Secure-Space Worker, the Job Scheduler and Execution Engine and the Execution Cluster, as well as running instances of the Encryption Manager and the Decryption Manager.

The **Secure-Space Worker** is the component residing at the Secure and Private Space that undertakes the responsibility of the executing the actual data analysis on the Secure and Private Space by interpreting the instructions provided by the Core Platform. The Secure-Space Worker is interacting with the Master Controller component towards the realisation of the Master / Worker paradigm that is adopted in the ICARUS platform architecture in order to locally execute the designed in the Core Platform data analysis jobs utilising the local services of the Secure and Private Space. Hence, the Secure-Space Worker is performing the following main operations based on the received instructions: a) transfer the selected encrypted datasets that will be utilised in the data analysis in the Secure Storage, b) decrypt the encrypted datasets with the help of the Decryption Manage, c) execute the data analysis job with the help of the Jobs Scheduler and Execution Engine, d) the encrypt the produced results with the help of the Encryption Manager and store them in the Secure Storage.

The **Jobs Scheduler and Execution Engine** undertakes the execution of the analytics jobs as provided by the Analytics and Visualisation Workbench by managing the execution on its complete lifecycle. In this sense, the Job Scheduler and Execution Engine performs the deployment of the Execution Cluster that will perform the data analysis in an isolated and secure environment with the help of the Resource Orchestrator. The Job Scheduler and Execution Engine initiates the immediate or scheduled execution on the Execution Cluster constantly monitoring the whole operation. Furthermore, it supports all the data handling aspect of the job execution by ensuring the access and storage of the required datasets or produced results by interacting with the Secure Storage. The Jobs Scheduler and Execution Engine performing all the cluster management operations by interacting with the Kubernetes in order to effectively monitor and allocate the required resources, manage the job execution and report the status back the Analytics and Visualisation Workbench.

The **Execution Cluster** is the actual cluster-computing framework of the platform that is deployed within the isolate environment spawned by Kubernetes and is managed by the Jobs Scheduler and Execution Engine. In this sense, the Execution Cluster is exploited by the Jobs Scheduler and Execution Engine in order to perform the data analysis within a secure and isolated private instance of the cluster-computing framework. Under the hood , the dominant Apache Spark framework is utilised, exploiting the rich set of features offered by Spark, the offered powerful processing engine that supports the execution of an extended list of data analysis algorithms that span from simple statistical analysis to more advance and complex machine learning and deep learning algorithms

Within the Secure and Private Space, a running instance of the **Encryption Manager** and the **Decryption Manager** are also deployed. As described above, the role of the Encryption Manager in this tier is to encrypt the results of the analysis before they are securely transmitted and stored in the Core platform. On the other hand, the role of the Decryption Manager in this tier is the decryption of the dataset on the data consumer side in order to be used in the data analysis in the same manner as it is performed for the download and decryption of a dataset within the On-Premise Environment.

## 3    Results

The ICARUS platform aims to provide a novel big data-enabled platform with sophisticated and valuable platform features and offerings for the aviation data value chain. The development of the ICARUS platform followed the design specifications that were presented in the previous section adopting an iterative approach of incremental releases based on the Agile development methodology. However, as the implementation of the ICARUS platform was recently completed no scientific results have yet been made available. However, the ICARUS platform will be thoroughly verified and evaluated by four core use cases of the aviation data value chain in order to ensure the added value of the implemented platform and showcase its unique innovation potential in the aviation industry. The following sections briefly present these use cases, focusing on the description of the context of each use case and the expected added value from the utilisation of the ICARUS platform.

### 3.1  Extra-aviation services in an integrated airport environment

Airports maintain a crucial role across the aviation industry as they constitute the indispensable connection to all the aviation-related activities, companies and organisations that are directly or indirectly linked to aviation. One crucial parameter of their operation is the capacity of the airport infrastructure (stands, gates vs planned aircraft arrivals) and its efficient planning. The main goal of the airport capacity planning is to ensure that the capacity of the airport infrastructure is sufficiently meeting the demand of the airports stakeholders even during busy periods when the demand increases due to unforeseen reasons (i.e. difficult weather conditions) or due to increased incoming or outgoing traffic (i.e. high season travel periods).

The scope of the specific use case is to facilitate the effective capacity enhancement decisions that will enable the sustained increase in throughput performance and will increase capacity in all weather conditions. In this use case, the variety of the descriptive and predictive analytics algorithms offered by the ICARUS platform will be exploited towards the following main axes of the airport capacity planning: a) the Capacity Modelling, b) the Airport Traffic Forecasting, c) the Flight Delay Prediction and d) the Position and Slot Allocation / Scheduling. Through the ICARUS platform, the expected added value is the enhancement of the Airport Airside Capacity with the resource usage optimisation of the airport airside infrastructure and the enhancement of the runway Operations Capacity. In the initial phase of the use case execution, the focus will be on the following aspects of the Capacity Modelling and Forecasting: i) the enhanced planning of flight schedules per season, ii) the optimised coordination of ground services and iii) the enhancement of the airport operation services.

### 3.2  Routes analysis for fuel consumption optimization and pollution awareness

Airlines are constantly addressing the challenge of reducing the fuel expenses in order to reduce their operational expenses but also to adhere the imposed regulations for reducing the environmental impact of the airline activities. However, this requires extensive and flexible analysis of various parameters in order to be in position to correctly assess the viability of route network extensions or modifications from an economic perspective that is dependent on the prediction of key operational metrics such as block time, block fuel and payload capacity.

The scope of the specific use case is to enable the optimised analysis of pollution data and aircraft emissions by performing advanced modelling of pollution data towards the prediction of aircraft performance in conjunction with the environmental impact. Furthermore, within this use case the analysis will be taken one step further by scaling this analysis from a single route into a massive route network on the second phase of the use case execution, by modelling aircraft payload capacity scenarios, taking also into consideration the weather conditions, towards the prediction of the aircraft performance in relation with a massive route network. For this specific use case, an external tool capable of performing route analysis, aircraft performance and economic investigations will be exploited in order to perform the required pre-processing calculations that will be used as input in the data analysis process that will conducted within the ICARUS platform that will allow the extraction of insights for aircraft fuel burn and carbon emissions for defined flight legs or a massive route network through the proper visualisations.

### 3.3   Aviation-related disease spreading

The global spreading of epidemics requires the design and utilisation of the advanced and complex mathematical models. The scope of this mathematical tools is to provide additional insights on the various parameters of the spreading of epidemic and most importantly the useful analysis and forecasting of the evolution of each epidemic towards the effective policy making from the corresponding health authorities. To this end, a model associated to meta-population has been developed in order to exploit the available real data as collected during any epidemic outbreak.

The scope of the specific use case to introduce the required non-incremental improvements this meta-population model by leveraging the aviation-related data, and especially airline traffic data, in order to further model the relationship between the human mobility and the spread of an infectious disease. The use case will firstly leverage the aviation-related data in order to further optimise the existing model, as well as to perform a both qualitative and quantitative assessment of the model utilising also historical and current epidemic forecasts. To this end, the use case will exploit the offering of the ICARUS platform for data exploration and data exchange, as well as data preparation in term of pre-processing, anonymisation and cleaning in order to produce the updated model version that will be holistically evaluated. On the second phase of the use execution, more passenger demographical data will be leveraged and combined with population-related and airline data in formulate the modelling framework that will enable the full coupling between human mobility and intra-population interactions.

### 3.4   Enhancing passenger experience with ICARUS data

A major ongoing challenge of all the service providers on the aviation industry is the enhancement of the passenger experience with the optimisation of the provided services and offering to the passenger from the moment they arrive at the airport till the moment that they arrive at their destination. Hence, the specific use case will demonstrator the added value obtained by the ICARUS platform towards this aim by focusing on two different axes, the prediction of on-board sales and the optimisation of the tray loading towards the reduce of cabin food waste and the increase of revenue and the prediction of profitable discounts and offers towards the increase of inflight sales.

The scope of the specific use case is to demonstrate the added value that can be obtained through the predictive algorithms and methods offered by the ICARUS platform for the prediction of the optimal loading weight of the duty-free and catering trays on board that can be exploited by both catering services and airlines prior to each flight in order to reduce cabin food waste and optimise their profits. Furthermore, the specific use case will demonstrate how the utilisation of the predictive algorithms and methods offered by the ICARUS platform can provide added value to airlines and catering service companies by predicting valuable discounts and offers on products and bundles that will increase in-flight sales while at the same time increase passenger experience.

## 4    Conclusions

The scope of the current paper is to introduce the ICARUS big data-enabled platform that aims provide a multi-sided platform that offers a novel aviation data and intelligence marketplace accompanied by a trusted and secure "sandboxed" analytics workspace. The ICARUS platform holistically handles the complete big data lifecycle from the data collection, data curation and data exploration to the data integration and data analysis of data originating from heterogeneous data sources with different velocity, variety and volume in a trusted and secure manner. The platform exploits methods such as big data analytics, deep learning, data enrichment, and blockchain powered data sharing, in order to properly address critical barriers for the adoption of Big Data in the aviation industry facilitating the design and execution of big data scenarios from the stakeholders of the aviation industry. In order to verify, validate and evaluate the ICARUS concept, approach and technical solution the four core representative use cases of the overall aviation's value chain, as briefly presented, will be performed.

## 5    Acknowledgement

ICARUS project is being funded by the European Commission under the Horizon 2020 Programme (Grant Agreement No 780792).